\documentclass[%
 aip,
 jmp,%
 amsmath,amssymb,
 reprint,%
]{revtex4-1}

\usepackage{graphicx}% Include figure files
\usepackage{dcolumn}% Align table columns on decimal point
\usepackage{bm}% bold math
\newcommand{\bA}{\mathbf{A}}
\newcommand{\bB}{\mathbf{B}}
\newcommand{\bC}{\mathbf{C}}

\newcommand{\bG}{\mathbf{G}}

\newcommand{\bR}{\mathbf{R}}

\newcommand{\br}{\mathbf{r}}
\newcommand{\brp}{\mathbf{r}'}
\newcommand{\brpp}{\mathbf{r}''}
\newcommand{\bkq}{\mathbf{k}+\mathbf{q}}

\newcommand{\bk}{\mathbf{k}}
\newcommand{\bq}{\mathbf{q}}
\newcommand{\bkp}{\mathbf{k}'}

\newcommand{\bkqp}{\mathbf{k}'+\mathbf{q}'}
\newcommand{\bzero}{\mathbf{0}}

\newcommand{\coul}{(\br, \brp)}
\newcommand{\coulr}{|\br - \brp|}

\newcommand{\ewalda}{|\br - \brp - \bA|}
\newcommand{\ewaldb}{|\br - \brp - \bB|}

\newcommand{\dr}{\int d\br}

\newcommand{\drdrp}{\int d\br d\brp}

\newcommand{\vmnbq}{$V_{mn\beta}^{\bk,\bq}$ }

\newcommand{\ibzvol}{\Omega_{BZ}/N^3}

\preprint{AIP/123-QED}

\begin{document}

\title{Density fitting in periodic systems: application to TDHF in diamond and oxides}

\author{C. H. Patterson}
\affiliation{School of Physics, Trinity College Dublin, Dublin 2, Ireland}
\email{Charles.Patterson@tcd.ie}

\date{\today} 

\begin{abstract}
A robust density fitting method for calculating Coulomb matrix elements over Bloch functions based on calculation of
two- and three-center matrix elements of the Ewald potential is described
and implemented in a Gaussian orbital basis in the Exciton code. The method is tested by comparing Coulomb and
exchange energies from density fitting to corresponding energies from SCF HF calculations for diamond, magnesium oxide and bulk Ne. 
Density fitting coefficients from the robust method are compared to coefficients from a variational method
applied to wave function orbital products in bulk Ne. Four center Coulomb matrix elements from density fitting are 
applied to time dependent Hartree-Fock (TDHF) calculations in diamond, magnesium oxide and anatase and rutile polytypes of titanium dioxide. 
Shifting virtual states downwards uniformly relative to occupied states and 
scaling the electron-hole attraction term in the TDHF Hamiltonian by 0.4 yields good agreement with either experiment and/or 
Bethe-Salpeter equation calculations. This approach mirrors similar 'scissors' adjustments of occupied and virtual states and 
introduction of a scaled electron-hole attraction term in some time dependent DFT calculations.
\end{abstract}


\maketitle

\section{Introduction\label{introduction}}

Excitons in condensed matter arise from a balance of electron-hole Coulombic attraction and electron-hole pair hopping.
The former is mediated by finite wavevector $\bq$ ladder matrix elements over four Bloch functions and, for $\bq \rightarrow \bzero$
excitons which couple to light, the latter is mediated by ring matrix elements. Computation of these matrix elements
is expensive and thus far has mainly been done in condensed phases using plane wave basis sets. Here we describe a method
for computing these matrix elements using a Gaussian orbital basis and apply them to time dependent Hartree-Fock (TDHF) calculations
in diamond, magnesium oxide and titanium dioxide.
When a local orbital basis
set is used for a periodic system, methods capable of treating the long range nature of the Coulomb interaction between local orbital
basis functions are essential. Furthermore, four center matrix elements over the local orbital basis must be transformed to
a Bloch orbital basis, which may also be expensive. Density fitting (DF) of products of wave function orbitals is a long established 
technique in both finite 
\cite{Whitten73,Dunlap79,Mintmire82a,Dunlap00,Jung05,Reine08,Pedersen09,Koster09,Geudtner12,Mejia_Rodriguez14} and periodic systems
\cite{Mintmire82,Rohlfing95,Maschio07,Usvyat07,Milko07,Maschio08,Varga08,Burow09,Dunlap10,Katouda10,Maschio11,Lorenz12,DelBen13,Francini14,Sun17a} 
which can account for the long range Coulomb interaction and reduces the time required for integral calculation and transformation.

DF factorizes calculation of four center matrix elements of the Coulomb potential into products of two center Coulomb and three centre 
Coulomb or overlap integrals, depending on metric choice. This reduces the transformation from the local orbital basis to the Bloch orbital 
basis from a four orbital to a two orbital problem. When this is combined with space group symmetries of pairs of k points 
($\bk$ and $\bkq$) in reciprocal space, high density k point meshes (up to 14 x 14 x 14 in this work) can be reached.
All calculations reported were done using the Exciton code \cite{Patterson10,Patterson19} which uses a local Gaussian orbital basis
for SCF Hartree Fock (HF) and time dependent Hartree Fock (TDHF) calculations on finite and periodic systems. The code
is also capable of performing $GW$ and Bethe-Salpeter equation (BSE) calculations in finite systems \cite{Patterson19}.

Here we report periodic TDHF calculations on diamond and three oxide compounds. First of all
the theory underlying the methods used is described, secondly the accuracy of the DF approach used is evaluated by comparing 
Coulomb and exchange energies derived from DF to the corresponding SCF energies. The method used is a robust
DF method in which the error in the difference between the fitted and true densities of products of pairs of orbitals is minimized
with a Coulomb metric. However, net charges associated with these products are not reproduced exactly. The method of Lagrange multipliers 
is used to constrain fitted densities to their exact values in a variation DF method. Differences in constrained and unconstrained 
fitting coefficients are evaluated for bulk $fcc$ Ne. In the following section, results of TDHF calculations for the systems just mentioned 
are reported. It is well known that differences in energy eigenvalues from DFT or HF calculations significantly under or overestimate 
quasi-particle band gaps in solids. This can be corrected in a $GW$ calculation. However, since no periodic $GW$ method is yet available 
for the Exciton code, an approach which has been successfully employed in TDDFT calculations \cite{Botti04} is adopted.
Virtual states are shifted down by a constant amount and electron-hole attraction matrix elements are scaled by a factor of 0.4 for both 
diamond and the oxides studied. This is similar to the approach of Botti \textit{et al.} who introduced a term $-\alpha/q^2$ into the 
TDDFT exchange-correlation kernel for a range of insulating materials \cite{Botti04}. Scaling of these matrix elements replaces 
screening by the static, inverse dielectric function \cite{Rohlfing00}.

The charge densities which are fitted in this work are products of crystal orbitals, which are Gaussian orbitals with 
Bloch translational symmetry. Our DF approach employs an auxiliary basis in which HF Bloch functions are expanded. An alternative approach
is to use the properties of spherical harmonics to reduce the number of orbital products \cite{Foerster08}.
Since individual Gaussian orbitals have limited extent,
the net charge associated with a particular orbital product is finite and is equal to a Fourier transformed overlap
matrix element for that pair of Gaussian orbitals. 

In general these charge distributions have nonzero charge and dipole moment.
The importance of treating charge distributions in periodic systems with zero monopole, dipole and quadrupole moment 
\cite{Harris75,Saunders92} has frequently been emphasized in work on DF in periodic systems \cite{Lorenz11,Francini14,Sun17a}. 
Here we adopt an approach to fitting these orbital products in which this is taken into consideration using a conventional 
Ewald approach \cite{Saunders92}. Ring or bubble matrix elements in the TDHF Hamiltonian are constructed from orbital products 
at single $\bk$ points and are fitted via unmodulated Ewald sums ($\bq$ = 0). Ladder matrix elements (i.e. the electron-hole attraction)
are constructed from orbital products at k points separated by wave vector $\bq$ and are fitted via modulated Ewald sums. 
In the former the $\bG$ = 0 term is omitted and agreement between DF and SCF Coulomb energies is of the order of 20-40 $\mu$H per atom.
In the latter the $\bq \rightarrow 0$ limit must be taken. When the TDHF Hamiltonian is set up for a periodic system, this
limit can only be reached by extrapolation.

Differences in Coulomb and exchange energies from DF and from SCF energies in this work are similar to those reported previously.
For example, Burow \textit{et al.} reported DF calculations of Coulomb energies in molecules, molecular crystals, graphite and 
diamond \cite{Burow09}. They found differences ranging from 4 to 37 $\mu$H per atom in finite, molecular systems and from 3 to 
51 $\mu$H per atom in periodic systems. Sun \textit{et al.} \cite{Sun17a} calculated the Coulomb and exchange energies of a 
H lattice with the cubic diamond structure. Using a Gaussian orbital auxiliary basis they reproduced the SCF Coulomb energy 
to within 100$\mu$H and the exchange energy to 1mH per atom. For Si in the diamond structure they reproduced the HF energy 
per cell to within 100 $\mu$H. Supplementing the auxiliary bases with plane waves allowed energies to be reproduced to much 
greater accuracy (nH for Si). Milko \textit{et al.} \cite{Milko07} reported differences in HF energies \textit{etc.} in
trans-polyacetylene (t-PA) and similar 1-D polymers of 60 $\mu$H per unit cell and minimal errors in the HF band gap less than 1 meV. 

Computation of Coulomb and exchange matrix elements using DF for post SCF methods such as M\"{o}ller-Plesset methods has been
reported by Katouda and Nagase \cite{Katouda10} for trans-polyacetylene using Gaussian and Poisson DF basis sets. They found 
correlation energy differences associated with DF as small as 30 $\mu$H and ranging up to 3 mH using mixed Gaussian 
and Poisson auxiliary basis sets. Maschio and coworkers reported DF local MP2 calculations \cite{Maschio07,Maschio08} which
have been applied to metal organic framework systems with over 100 atoms per unit cell \cite{Maschio11}.
Lorenz \textit{et al.} performed configuration interaction-singles calculations for wide gap semiconductors and oxides 
using a local Wannier orbital approach. They found HF/TDHF band gaps of 16.29/11.94 eV for MgO and 14.65/11.72 for diamond C \cite{Lorenz12}. 
More recently, Mackrodt \textit{et al.} reported calculations of the optical reflectivity of $\alpha$-Al$_2$O$_3$
using a finite frequency, coupled perturbed B3LYP method \cite{Mackrodt20} in a Gaussian local orbital basis.

\section{Theory\label{theory}}

\subsection{Robust Density Fitting\label{robust_DF}}

Following Dunlap \cite{Dunlap00,Dunlap10}, robust fitting of the charge densities in a Coulomb integral,

\begin{equation}
\label{Eqn1}
\left<\rho(\br)|\rho(\brp)\right> = \drdrp \rho(\br) v\coul \rho(\brp).
\end{equation}

\noindent
results from minimization of the quadratic error, 

\begin{equation}
\label{Eqn2}
\left<\Delta \rho(\br)|\Delta \rho(\brp)\right> = \drdrp \Delta \rho(\br) w(\br - \brp) \Delta \rho(\brp).
\end{equation}

\noindent
where $v\coul$ is the Coulomb potential, $\Delta \rho(\br) = \rho(\br) - \tilde{\rho}(\br)$ is the difference in 
true and fitted densities and $w(\br - \brp) = v\coul$ is the Coulomb metric.

Electron-hole hopping and electron-hole attraction matrix elements needed for TDHF calculations are,

\begin{equation}
\label{Eqn3}
\left <\Psi_{v\bk}^*(\br) \Psi_{c\bk}(\br) | \Psi_{c'\bkq}^*(\brp) \Psi_{v'\bkq}(\brp)\right>
\end{equation}

and 

\begin{equation}
\label{Eqn4}
\left < \Psi_{v\bk}^*(\br) \Psi_{v'\bkq}(\br) | \Psi_{c'\bkq}^*(\brp) \Psi_{c \bk}(\brp)\right>,
\end{equation}

\noindent
respectively. $\Psi_{v\bk}(\br)$ and $\Psi_{c \bkq}(\br)$ are valence and conduction band states at wave vectors $\bk$ and $\bkq$.
Bloch functions, $\Psi_{v\bk}(\br)$, are expanded as linear combinations of phase modulated local orbitals, $\phi_m(\br - \bR)e^{i\bk.\bR}$, 
with lattice translation vector, $\bR$, 

\begin{equation}
\label{Eqn5}
\Psi_{v\bk}(\br) =  \sum_{m,\bR} d_m^{v\bk} \phi_m(\br - \bR) e^{i\bk.\bR},
\end{equation}

\noindent
which are referred to as crystal orbitals (CO) with wave vector, $\bk$,
and expansion coefficients, $d_m^{v\bk}$. Expansion of Bloch function products in Eq. \ref{Eqn3} and \ref{Eqn4} leads to CO products,

\begin{equation}
\label{Eqn6}
\rho^{\bk,\bq}_{mn}(\br) = \sum_{\bA,\bB}\phi_m^*(\br-\bA)\phi_n(\br-\bB) e^{-i\bk.\bA + i(\bkq).\bB},
%\rho^{\bq}_{mn}(\br) = \sum_{\bA,\bB}\phi_m^*(\br-\bA)\phi_n(\br-\bB) e^{-i\bk.\bA + i(\bkq).\bB},
\end{equation}

\noindent
the charge densities which are fitted in this work. %The dependence of these densities on $\bk$ is suppressed for compactness of notation.
An auxiliary  basis of the form,

\begin{equation}
\label{Eqn7}
\chi_\alpha^{\bq}(\br) =  \sum_{\bR} \chi_\alpha(\br - \bR) e^{i\bq.\bR}
\end{equation}

\noindent
where the auxiliary basis, $\chi_\alpha^{\bq}(\br)$, is a more extensive Gaussian orbital basis set than the CO basis in Eq. \ref{Eqn5}
is used to expand the Bloch functions.

The condition that auxiliary function expansion coefficients, $c_{\alpha}^{\bk,\bq}$, minimize the error,

\begin{equation}
\label{Eqn8}
\left< \rho^{\bk,\bq}_{mn}(\br) - c_{\alpha}^{\bk,\bq} \chi_{\alpha}^{\bq}(\br) | \rho^{*\bk,\bq}_{rs}(\brp) - c_{\beta}^{*\bk,\bq} \chi_{\beta}^{*\bq}(\brp) \right>
%\left< \rho^{\bq}_{mn}(\br) - c_{\alpha}^{\bq} \chi_{\alpha}^{\bq}(\br) | \rho^{*\bq}_{rs}(\brp) - c_{\beta}^{*\bq} \chi_{\beta}^{*\bq}(\brp) \right>
\end{equation}

\noindent
in the Coulomb integral $\left<\rho^{\bk,\bq}_{mn}(\br) | \rho^{*\bk,\bq}_{rs}(\brp)\right>$ is,
%in the Coulomb integral $\left<\rho^{\bq}_{mn}(\br) | \rho^{*\bq}_{rs}(\brp)\right>$ is,

\begin{equation}
\label{Eqn9}
c_{\alpha}^{\bk,\bq} \left< \chi_{\alpha}^{\bq}(\br) | \chi_{\beta}^{*\bq}(\brp) \right> = \left< \rho_{mn}^{\bk,\bq}(\br) | \chi_{\beta}^{*\bq}(\brp) \right>
%c_{\alpha}^{\bq} \left< \chi_{\alpha}^{\bq}(\br) | \chi_{\beta}^{*\bq}(\brp) \right> = \left< \rho_{mn}^{\bq}(\br) | \chi_{\beta}^{*\bq}(\brp) \right>
\end{equation}

\noindent
Manipulation of the lattice sums in Eq. \ref{Eqn6}, \ref{Eqn7} and Eq. \ref{Eqn8} using lattice translational invariance turns Eq. \ref{Eqn9}
into the following lattice modulated Ewald sums \cite{Ewald21,Saunders92} (Appendix A),

\begin{widetext}

\begin{equation}
\label{Eqn10}
%c_{\alpha}^{\bq}\sum_{\bA} \frac{\chi_{\alpha}(\br) \chi_{\beta}^{*}(\brp) } {\ewalda}  e^{-i\bq.\bA} = 
c_{\alpha}^{\bk,\bq}\sum_{\bA} \drdrp \frac{\chi_{\alpha}(\br) \chi_{\beta}^{*}(\brp) } {\ewalda}  e^{-i\bq.\bA} = 
\sum_{\bB,\bC} \drdrp \frac{ \phi_{m}^{*}(\br) \phi_{n}(\br - \bC) \chi_{\beta}^{*}(\brp) } {\ewaldb} e^{i(\bkq).\bC} e^{-i\bq.\bB},
\end{equation}

\noindent
where the periodic Ewald potential is shown explicitly for clarity. The potential,

\begin{equation}
\label{Eqn11}
\sum_{\bA} \frac{e^{i\bq.\bA}}{\ewalda} = \sum_{\bG} \frac{4\pi}{\Omega} \frac{e^{-\frac{|\bq + \bG|^2}{4\gamma}}}{|\bq + \bG|^2} 
e^{i(\bq + \bG).(\br - \brp)} + \sum_{\bA} \frac{\text{erfc}(\gamma^{1/2}|\br - \brp - \bA|)}{|\br - \brp - \bA|} e^{i\bq.\bA},
\end{equation}

\end{widetext}

\noindent
is a lattice-modulated generalization of the familiar Ewald potential \cite{Born88}, where $\gamma$ is the splitting parameter
in Ewald's method and $\Omega$ is the unit cell volume. Defining,

\begin{equation}
\label{Eqn12}
V_{\alpha\beta}^{\bq} = \sum_{\bA} \drdrp \frac{\chi_{\alpha}(\br) \chi_{\beta}^{*}(\brp) } {\ewalda}  e^{-i\bq.\bA} 
\end{equation}

\noindent
\begin{equation}
\label{Eqn13}
V_{mn\beta}^{\bk,\bq} = \sum_{\bB,\bC} \drdrp \frac{ \phi_{m}^{*}(\br) \phi_{n}(\br - \bC) \chi_{\beta}^{*}(\brp) } {\ewaldb} e^{i(\bkq).\bC} e^{-i\bq.\bB}
%V_{mn\beta}^{\bq} = \sum_{\bB,\bC} \drdrp \frac{ \phi_{m}^{*}(\br) \phi_{n}(\br - \bC) \chi_{\beta}^{*}(\brp) } {\ewaldb} e^{i(\bkq).\bC} e^{-i\bq.\bB}
\end{equation}

\noindent
the equation which yields the coefficients which minimize the error in Eq. \ref{Eqn8} becomes 
$c_{\alpha}^{\bk,\bq} V_{\alpha\beta}^{\bq} = V_{mn\beta}^{\bk,\bq}$ and the expansion coefficients are,
%$c_{\alpha}^{\bq} V_{\alpha\beta}^{\bq} = V_{mn\beta}^{\bq}$ and the expansion coefficients are,

\begin{equation}
\label{Eqn14}
c_{\alpha}^{\bk,\bq} = V_{mn\beta}^{\bk,\bq} V_{\beta\alpha}^{\bq-1},
%c_{\alpha}^{\bq} = V_{mn\beta}^{\bq} V_{\beta\alpha}^{\bq-1},
\end{equation}

\noindent
where $V_{\alpha\beta}^{\bq-1}$ is the matrix inverse of $V_{\alpha\beta}^{\bq}$.
Substitution of the expansion of the density, $\rho^{\bk,\bq}_{mn}(\br) = c_{\alpha}^{\bk,\bq} \chi_{\alpha}^{\bq}(\br)$, yields
%Substitution of the expansion of the density, $\rho^{\bq}_{mn}(\br) = c_{\alpha}^{\bq} \chi_{\alpha}^{\bq}(\br)$, yields

\begin{equation}
\label{Eqn15}
\drdrp \frac{ \rho^{\bk,\bq}_{mn}(\br) \rho^{\bk,\bq*}_{rs}(\brp) }{\coulr} \approx c_{\alpha}^{\bk,\bq} \drdrp \frac{ \chi_{\alpha}^{\bq}(\br) \rho^{*\bk,\bq}_{rs}(\brp) }{\coulr}
%\drdrp \frac{ \rho^{\bq}_{mn}(\br) \rho^{\bq*}_{rs}(\brp) }{\coulr} \approx c_{\alpha}^{\bq} \drdrp \frac{ \chi_{\alpha}^{\bq}(\br) \rho^{*\bq}_{rs}(\brp) }{\coulr}
\end{equation}

\begin{equation}
\label{Eqn16}
\drdrp \frac{ \rho^{\bk,\bq}_{mn}(\br) \rho^{*\bk,\bq}_{rs}(\brp) }{\coulr} \approx  V_{mn\beta}^{\bk,\bq} V_{\beta\alpha}^{\bq-1} V_{\beta rs}^{*\bk,\bq}
%\drdrp \frac{ \rho^{\bq}_{mn}(\br) \rho^{\bq*}_{rs}(\brp) }{\coulr} \approx  V_{mn\beta}^{\bq} V_{\beta\alpha}^{\bq-1} V_{\beta rs}^{\bq*}
\end{equation}

\subsection{Variational Density Fitting\label{variational_DF}}

For some applications, such as use of DF in self-consistent field calculations, it is desirable or essential
that fitted densities not only minimize errors in Coulomb and exchange energies, but also conserve charge. The robust fit method obtained by 
minimizing the error in Eq. \ref{Eqn9} is not variational \cite{Dunlap79,Dunlap00} 
in that while errors in the electrostatic energy in Eq. \ref{Eqn1} are minimized, the fitted charge densities are not constrained to
equal the densities being fitted. The constraint that the integrated charge densities $Q^{\bk,\bq}_{mn} = \dr \rho^{\bk,\bq}_{mn}(\br)$ and 
$\overline{Q}^{\bk,\bq}_{mn} = \overline{c}^{\bk,\bq}_{\alpha} \dr \chi^{\bq}_\alpha(\br)$ be equal is imposed by minimizing the functional,

\begin{widetext}

\begin{equation}
\label{Eqn17}
\left<\Delta \rho^{\bk,\bq}_{mn}(\br)|\Delta \rho^{*\bk,\bq}_{rs}(\brp)\right> - \lambda^{\bk,\bq} \dr (\rho^{\bk,\bq}_{mn}(\br) - \overline{c}^{\bk,\bq}_\alpha \chi_\alpha(\br)) - \mu^{*\bk,\bq} \dr (\rho^{*\bk,\bq}_{rs}(\br) - \overline{c}^{*\bk,\bq}_{\beta}\chi_{\beta}^{*}(\br))
%\left<\Delta \rho^{\bk,\bq}_{mn}(\br)|\Delta \rho^{*\bk,\bq}_{rs}(\brp)\right> - \lambda^{\bk,\bq} \dr (\rho^{\bk,\bq}_{mn}(\br) - \overline{c}^{\bk,\bq}_\alpha \chi_\alpha^{\bq}(\br)) - \mu^{*\bk,\bq} \dr (\rho^{*\bk,\bq}_{rs}(\br) - \overline{c}^{*\bk,\bq}_{\beta}\chi_{\beta}^\bq(\br))
\end{equation}

\end{widetext}

\noindent
Where $\lambda$ and $\mu$ are Lagrange multipliers. The constraint,

\begin{equation}
\label{Eqn18}
Q^{\bk,\bq}_{mn} = \overline{Q}^{\bk,\bq}_{mn},
%\dr (\rho^{\bq}_{mn}(\br) - \overline{c}^{\bq}_{\alpha}\chi^{\bq}_\alpha(\br)) = 0,
%\dr (\rho^{\bk,\bq}_{mn}(\br) - \overline{c}^{\bk,\bq}_{\alpha}\chi^{\bq}_\alpha(\br)) = 0,
\end{equation}

\noindent
reduces to,

\begin{equation}
\label{Eqn19}
\sum_{\bB'} \dr \phi_m^*(\br)\phi_n(\br-\bB') e^{i(\bk+\bq).\bB'} = \overline{c}^{\bk,\bq}_\alpha \dr \chi_\alpha(\br),
%\sum_{\bB'} \dr \phi_m^*(\br)\phi_n(\br-\bB') e^{i\bk.\bB'} = \overline{c}^{\bk,\bq}_\alpha \dr \chi_\alpha(\br),
\end{equation}

\noindent
using lattice translational invariance (Appendix B). The term on the left in Eq. \ref{Eqn19}
is the Fourier transform of the overlap matrix at wave vector $\bkq$, $S_{mn}^{\bkq}$. The bar on the coefficient 
$\overline{c}_\alpha^{\bk,\bq}$ indicates a constrained coefficient. The resulting linear equations, which replace Eq. \ref{Eqn14}, are,

\begin{equation}
\label{Eqn20}
\left( \begin{array}{cc}
V_{\beta\alpha}^{\bq} & \left<\beta \right>^* \\
\left < \alpha \right>  & 0 \end{array} \right)
\left( \begin{array}{c}
\overline{c}_\alpha^{\bk,\bq} \\ \mu^{*\bk,\bq} \end{array} \right) = 
\left( \begin{array}{c}
V_{mn\beta}^{\bk,\bq} \\ S_{mn}^{\bkq} \end{array} \right),
\end{equation}

\noindent
where $\left< \alpha \right> = \dr \chi_\alpha(\br)$ is nonzero for $l = 0$ auxiliary functions and zero otherwise.
For an auxiliary basis function with a single, normalized Gaussian this equals $(2\pi/a)^{3/4}$, where $a$ is the Gaussian exponent.
Eq. \ref{Eqn20} must be solved for each unique $\bk$ and $\bkq$ pair, however, decomposition of the matrix
on the left before solution of these equations needs to be performed only once for each unique $\bq$ vector.
Here we investigate how well the unconstrained, robust density fitting of Eq. \ref{Eqn14} satisfies charge conservation
for bulk \textit{fcc} Ne by comparing robust and variational DF expansion coefficients. All other calculations reported here used 
robust, non-variational fitting.

\subsection{Exchange and Coulomb energies\label{exch_coul}}

Comparison of Coulomb and exchange energies from SCF and DF calculations presents a useful means of evaluating errors in 
density fitted electrostatic energies in periodic systems. Coulomb and exchange energies, $E_C$ and $E_x$, are obtained 
by replacing conduction band states by valence states in Eq. \ref{Eqn3} and \ref{Eqn4} and summing over wave vectors, $\bk$ and $\bkq$, 

\begin{equation}
\label{Eqn21}
E_C = \frac{1}{2N_\bk N_\bq} \sum_{v,v',\bk,\bq} \left <\Psi_{v\bk}^*(\br) \Psi_{v\bk}(\br) | \Psi_{v'\bkq}^*(\brp) \Psi_{v'\bkq}(\brp)\right>
\end{equation}

and 

\begin{equation}
\label{Eqn22}
E_x = -\frac{1}{4N_\bk N_\bq}\sum_{v,v',\bk,\bq} \left<\Psi_{v\bk}^*(\br) \Psi_{v'\bkq}(\br) | \Psi_{v'\bkq}^*(\brp) \Psi_{v \bk}(\brp)\right>.
\end{equation}

In order to transform the density fitted matrix elements over the local orbital basis in Eq. \ref{Eqn16} into matrix elements over
Bloch functions,

\begin{equation}
\label{Eqn23}
 V_{\alpha v\bk v'\bkq}^{\bq} = V_{\alpha mn}^{\bk,\bq} d_m^{v\bk}d_n^{v'\bkq},
\end{equation}

\noindent
where $d_m^{v\bk}$ and $d_n^{v'\bkq}$ are Bloch function expansion coefficients and the superscript $\bq$ indicates the $\bq$
vector used in $V_{\alpha mn}^{\bk,\bq}$. Coulomb and exchange energies in Eq. \ref{Eqn21} 
and \ref{Eqn22} become,

\begin{equation}
\label{Eqn24}
E_C \approx \frac{1}{2N_\bk N_\bq} \sum_{v,v',\bk,\bq} V_{v\bk v\bk \alpha}^{\bzero} V_{\alpha\beta}^{\bzero,-1} \
V_{\beta v'\bkq v'\bkq}^{*\bzero}
\end{equation}

and

\begin{equation}
\label{Eqn25}
E_x \approx -\frac{1}{4N_\bk N_\bq} \sum_{v,v',\bk,\bq} V_{v\bk v'\bkq \alpha}^{\bq} V_{\alpha\beta}^{\bq,-1} V_{\beta v'\bkq v\bk}^{*\bq}
\end{equation}

\noindent
Differences in total Coulomb and exchange energies obtained using these expressions and from SCF calculations used to generate 
Bloch functions in Eq. \ref{Eqn24} and \ref{Eqn25} are reported below. These are in the 25 - 50$\mu$H per atom range for Coulomb energies 
and 1mH per atom range for exchange energies extrapolated to infinite $\bq$ sampling density.

\subsection{Small $\bq$ limit\label{small_q}}

The divergent nature of the $V_{\alpha\beta}^{\bq}$ and $V_{mn\beta}^{\bk,\bq}$ matrix elements in Eq. \ref{Eqn12}
and \ref{Eqn13} around $\bq = \bzero$ requires special attention. The contribution to the Coulomb energy at $\bq = \bzero$ is
straightforward. The first term on the right in Eq. \ref{Eqn11} with $\bG = \bzero$ is replaced by $-\pi/\gamma\Omega$
in a 3-D periodic system \cite{Born88}. %$\Omega$ is the unit cell
%volume and $\gamma$ is the Ewald parameter used to split interactions over real and reciprocal space. 
Choosing $\gamma \rightarrow \infty$ makes the real space sum on the right tend to zero, leaving the Fourier expansion of the
Coulomb potential on the right. 

Expansion of the complex exponential in this term for small $\bq$ and $\bG = \bzero$ as $e^{i\bq.(\br-\brp)} \approx 1 + i\bq.(\br-\brp)$
allows small $\bq$ contributions for various interactions to be determined. Valence and conduction states in
$\Psi^*_{v\bk}$ and $\Psi_{c\bk}$ or $\Psi^*_{v'\bkq}$ and $\Psi_{c'\bkq}$ in Eq. \ref{Eqn3} (ring diagrams) are orthogonal at small
$\bq$. Hence the leading contribution in this case is from $i\bq.(\br-\brp)$. This term (of order meV) leads to splitting of longitudinal
and transverse excitons \cite{Rohlfing00} and is omitted in this work. For valence states $\Psi^*_{v\bk}$ and $\Psi_{v'\bkq}$
which occur in the exchange energy (Eq. \ref{Eqn22}) and electron-hole attraction (ladder diagrams, Eq. \ref{Eqn4})
and conduction states $\Psi^*_{c'\bkq}$ and $\Psi_{c\bk}$ in ladder diagrams, identical states ($v = v'$ and $c = c'$) have unit
overlap as $\bq \rightarrow \bzero$ while distinct states are orthogonal. The former have a contribution from the leading term
in the expansion of $e^{i\bq.(\br-\brp)}$ and the latter in the $i\bq.(\br-\brp)$ term. 

The $\bq \rightarrow 0$ limit of the exchange energy and the electron-hole attraction term in the TDHF Hamiltonian (Eq. \ref{Eqn4}) 
is obtained by assuming that matrix elements around $\bq = \bzero$ vary slowly with $\bq$ and may be approximated by the average
value of the analytically integrated Coulomb potential around $\bq = \bzero$. Averaging $4 \pi / \Omega q^2$ over a sphere with 
volume $\ibzvol$ yields,

\noindent
\begin{equation}
\label{Eqn26}
\left< \frac{4\pi} {\Omega} \frac{1} {|\bq|^2}\right> = 4\left( \frac{3 N^2}{4\pi\Omega} \right)^{\frac{1}{3}}.
\end{equation}

\noindent
$N^3$ is the number of $\bq$ points in a regular Monkhorst-Pack net \cite{Monkhorst76}. This energy is used for each
state where $v = v'$ and $\bq \rightarrow \bzero$ in Eq. \ref{Eqn22}.

There are methods for calculating the total exchange energy more accurately than simply by sampling a divergent function of $\bq$.
Gygi and Baldereschi \cite{Gygi86} introduced a divergent, periodic function which is added and subtracted from
the divergent small $\bq$ limit of the Coulomb potential to produce a smoothly varying potential which is integrated
numerically and a divergent, periodic term which is integrated analytically over the Brillouin zone.
Sorouri \textit{et al.} \cite{Sorouri06}, Carrier \textit{et al.} \cite{Carrier07} and Duchemin and Gygi \cite{Duchemin10} 
later adopted a simpler auxiliary function,

\begin{equation}
\label{Eqn27}
F(\mathbf{k}) = \sum_\mathbf{G} \frac{e^{-\alpha|\mathbf{k}-\mathbf{G}|^2}}{|\mathbf{k} - \mathbf{G}|^2}
\end{equation}

\noindent
which may be applied to all crystal systems. Spencer and Alavi \cite{Spencer08} adopted a function which has a spherical 
real space cutoff. %The approaches were compared by Holzwarth and Xu \cite{Holzwarth11}. 

Here, however, the method of calculating TDHF spectra requires sampling of the electron-hole Hamiltonian on a regular 
Monkhorst-Pack net. The sampling method described above is used to calculate the contribution to the exchange energy and the
electron-hole Hamiltonian. Calculation of the exchange energy for a series of nets of increasing k point density allows the
exchange energy to be extrapolated to infinite density. This approach is used in results reported in Section \ref{results}.
Extrapolated exchange energies are compared to the exchange energy from the SCF calculation, which generated the Bloch functions
used in density fitted integrals. 

The exchange energy in these SCF calculations is obtained from four-center, real-space integrals and the real-space density matrix,

\begin{widetext}

\begin{equation}
\label{Eqn28}
E_x^{SCF} = -\frac{1}{4} \drdrp \phi^*_i(\br) \phi_j(\br-\bA) \frac{1}{|\br-\brp|} \phi_k(\brp-\bB) \phi^*_l(\brp-\bC) P^{\bC\bA}_{jl} P^{\bB}_{ik},
\end{equation}

\end{widetext}

\noindent
$P^{\bC\bA}_{jl}$ is the $jl$ element of the real-space density matrix at lattice vector $\bC - \bA$. This approach relies on
the convergence of the real-space density matrix with lattice vector range, $|\bC\bA|$ and $|\bB|$. The real-space density matrix 
is exponentially localized for a gapped material and the localization length reduces with increase of the band gap. In this work only
wide gap materials are studied and the SCF exchange energy is well converged for these systems. 

It is worth noting that the Ewald method of calculating exchange energies (Eq. \ref{Eqn22}) does not rely on convergence of the 
density matrix in real space because the Ewald method sums interactions to infinite range. It may therefore be a superior 
method of calculating the HF exchange operator
in SCF calculations on narrow gap or conducting systems. Indeed, products of Bloch functions in matrix elements in many-body
calculations are delocalized over all space, and therefore calculation of converged electron-hole attraction matrix elements 
in crystalline systems is not possible using real-space four-center integrals of the kind in Eq. \ref{Eqn28}. An Ewald approach 
such as that used here is essential. Casting the Coulomb interaction into reciprocal space in Eq. \ref{Eqn11} by choosing 
$\gamma \rightarrow \infty$ has the disadvantage that a very large number of $\bG$ vectors must be used in order to obtain 
convergence for even relatively low Gaussian exponents, (of order 1, say). Hence, the mixed real and reciprocal space method
advocated here has a number of important advantages.

\subsection{Time dependent Hartree-Fock method\label{TDHF}}

The TDHF equations are usually expressed as the following generalized eigenvalue problem,

\begin{equation}
\label{Eqn29}
\left( \begin{array}{cc}
A & B \\
-B^* & -A^* \end{array} \right)
\left( \begin{array}{c}
X \\ Y \end{array} \right) = 
\Lambda
\left( \begin{array}{c}
X \\ Y \end{array} \right). 
\end{equation}

\noindent
Here the Tamm-Dancoff approximation (TDA) to the TDHF equations is used throughout in which the off-diagonal $B$ blocks are omitted.
In this case the TDHF problem reduces to a standard eigenvalue problem with eigenvectors $X$ and eigenvalues $\Lambda$.
The matrix elements which appear in the $A$ block of the TDHF Hamiltonian matrix are given in Eq. \ref{Eqn3} and \ref{Eqn4}.
The $A$ matrix also contains differences in single particle eigenvalues on the diagonal. Results are reported in Section \ref{results}
where either differences in HF eigenvalues are used or where these differences are reduced by a constant 'scissors' shift of around 10 eV.

\section{Computational Methods\label{methods}}

The implementation of this method in the Exciton code is described in this section. Firstly, the method of calculation
of finite-$\bq$ matrix elements which appear in ladder diagrams is given, followed by zero-$\bq$ matrix elements
which appear in ring diagrams. Diagonalization of the $A$ matrix in Eq. \ref{Eqn29} is performed using the $pzheevx$
routine in Scalapack \cite{Scalapack}. Scalapack uses a block-cyclic matrix distribution. Here the row and column blocksize in
this distribution is chosen to be the number of transitions per k-point, i.e. the number of occupied times the number of virtual states
per k point in the active space. This means that the matrix to be diagonalized is 
split over cores by $\bq$ point and all matrix elements corresponding to a particular ($\bk,\bkq$) pair are sent to the same core.

(1) HF-SCF calculations are performed using a set of unique k-points in an $N$ x $N$ x $N$ Monkhorst-Pack net.
Wave functions at symmetry equivalent k-points may be generated by rotation of these unique k-points using one known symmetry operator
for each equivalent k-point. Phases of wave functions at unique and symmetry equivalent k-points are therefore unique for each 
wave function and k-point. 

(2) Lists of unique ($\bk,\bkq$) pairs are generated using the little group of $\bq$, whose symmetry operations leave $\bq$ invariant. 
A set of symmetry-unique $\bq$ points in the Brillouin zone is identified, then $\bk$ points which are unique under the little group 
for that $\bq$ vector are identified. Finally, all symmetry equivalent ($\bk,\bkq$) pairs are generated using the full set of
space group operations. Unique instances of these pairs are stored along with the symmetry operators which generated them 
from a given unique ($\bk,\bkq$) pair.

(3) Matrix elements over atomic orbitals in Eq. \ref{Eqn12} and \ref{Eqn13} for unique ($\bk,\bkq$) pairs are calculated. These are
transformed to matrix elements over Bloch functions at these points by multiplying in the $d_m^{v\bk}$ expansion coefficients 
(Eq. \ref{Eqn5}).

(4) Computation of matrix elements \vmnbq is distributed over cores by unique $\bq$ point. \vmnbq matrix elements for unique 
($\bk,\bkq$) pairs are computed and these are rotated to all symmetry equivalent equivalent ($\bkp,\bkqp$) pairs. Each wave function 
in each ($\bk,\bkq$) pair must be rotated twice; once to a unique ($\bk,\bkq$) pair and once more from there to an equivalent ($\bkp,\bkqp$) 
pair. 

Two consequences of this are that (i) wave functions generated by products of symmetry operations may differ by a phase factor from
the definitive phase in (1) and (ii) degenerate wave functions generated by rotation by the single, unique operator in (1) 
may be mixed when rotated twice. Both (i) and (ii) are accounted for by obtaining the overlap matrix of wave functions rotated by the single, 
unique operator with those rotated twice. Columns of this overlap matrix yield the linear combination of matrix elements 
which must be used to ensure the uniqueness of each symmetry equivalent wave function product at $\bkp$ and $\bkqp$. 

(5) Once a set of \vmnbq matrix elements and its symmetry equivalents have been computed, they are distributed to cores
according to the Scalapack block-cyclic distribution mentioned above using an MPI-Send blocking send.

(6) Calculation of $\bq = \bzero$ matrix elements which occur in ring diagrams is simplified as wave functions at only one $\bk$ point
occur in each matrix element as $\bq$ = 0. Matrix elements calculated at the unique set of $\bk$ points mentioned in (1)
can therefore be used to generate all equivalent ($\bkp,\bkp$) pairs in ring diagrams.
 
\section{Basis sets\label{basis}}

The wave function basis sets used in ths work are the DEF2-TZVP basis sets of Weigend and Ahlrichs \cite{Weigend05}, modified for the 
solid state by removing diffuse basis functions or increasing their exponents and the auxiliary basis sets are the DEF2-TZVP-RIFIT basis
sets \cite{Weigend98}. Basis functions with angular momenta with $l$ values greater than $l$ = 4 ($h$ functions and higher) 
were omitted. The DEF2-SVP, -TZVP and -QZVP basis sets for Ne used in Section \ref{conservation} were used without modification. 
Modifications to DEF2-TZVP basis sets for C, O, Mg and Ti are given in the Supplementary Information.

\section{Results\label{results}}

In this Section differences in Coulomb and exchange energies obtained from SCF calculations and by DF are 
compared for diamond, magnesium oxide and bulk \textit{fcc} Ne. Fitted charges, $Q_{mn}^{\bk,\bq}$ and $\overline{Q}_{mn}^{\bk,\bq}$ 
(Eq. \ref{Eqn18}), obtained by solving Eq. \ref{Eqn14} and \ref{Eqn20} are computed for bulk \textit{fcc} Ne. The latter 
charges are equal to their exact numerical values, $S_{mn}^{\bkq}$, owing to the constraints imposed in Eq. \ref{Eqn20}. 
Finally, results of TDHF calculations for these systems and two polymorphs of TiO$_2$ are presented and analyzed.

\subsection{Fitted density Coulomb and exchange energies}

\begin{figure}[htp]
\includegraphics[width=10cm,angle=-90]{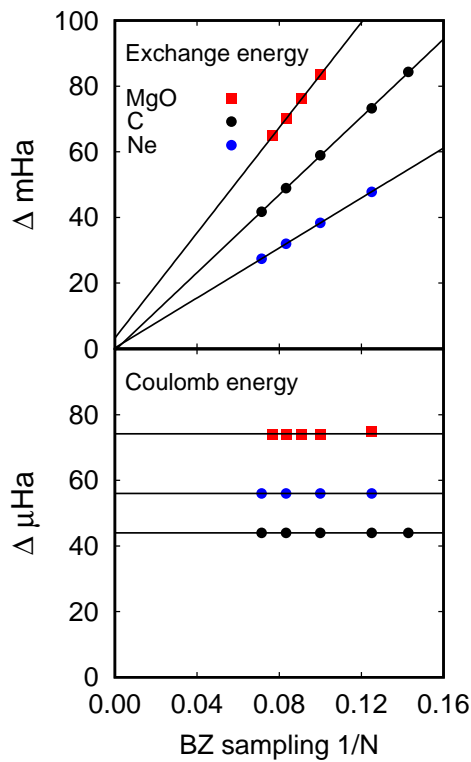}
\caption{(Color online) Difference in Coulomb and exchange energies from SCF and DF in bulk Ne, MgO and diamond in mHa.}
\label{fig1}
\end{figure}

Differences in Coulomb and exchange energies obtained from SCF calculations and by DF are shown in Fig. \ref{fig1}
as a function of Brillouin zone sampling frequency, 1/$N$. Differences in Coulomb energies are expected to be roughly independent
of sampling frequency (Section \ref{small_q}) as they depend only on matrix elements at $\bq = \bzero$. Fig. \ref{fig1} shows 
that differences in exchange energies from SCF calculations and DF scale with sampling frequency. Extrapolation of the DF exchange 
energy is needed for comparison to the SCF exchange energy. 

SCF and DF energy differences in Coulomb energies obtained from SCF calculations and DF are 45, 55 and 56 $\mu$H, respectively,
for diamond, MgO (around 20-25 $\mu$H per atom) and bulk Ne. These values are comparable to those obtained for molecular density fitting 
\cite{Reine08} using  a series of attenuated Coulomb metrics.
Extrapolated differences in SCF and DF exchange energies are -0.52 mHa for diamond, 3.26 mHa for MgO and 0.25 mHa for bulk Ne.
%When a TZVP basis and corresponding auxiliary basis are used unmodified for either the Ne atom or bulk, $fcc$ Ne, there is a 
%relatively large error in both the atomic and condensed phase Coulomb energy of 1.106 mH and 1.264 mH, respectively. However, 
%the 2$s$ valence region with an important exponent at 20.3 which produces a 2$s$2$s$ product 
%exponent at 40.6 is only fitted using exponents at 29.5 and 141.1. Introducing an additional $s$ Gaussian to the auxilairy basis
%with an exponent of 64.0 reduces the error in the Coulomb energy to 87 $\mu$H in the atom and 56$\mu$H in the condensed phase.
%The difference in SCF and extrapolated DF exchange energies in the condensed phase is reduced from 0.94 mH to 0.28 mH by introducing this
%additional auxiliary basis function.

\subsection{Charge conservation in fitted densities\label{conservation}}

\begin{figure}[hbp]
\includegraphics[width=7cm,angle=0]{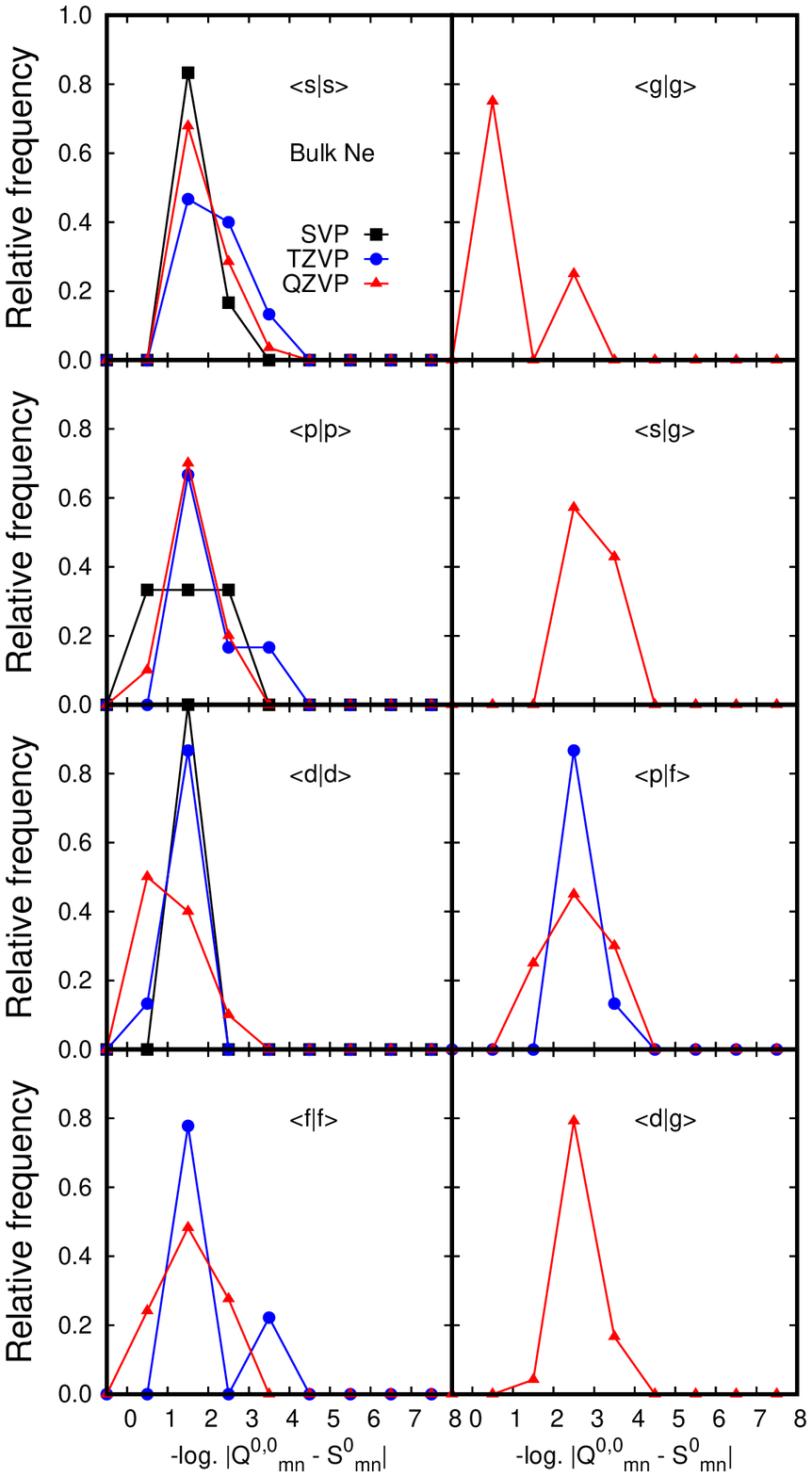}
\caption{(Color online) Distribution of differences in fitted charges, $Q_{mn}^{\bzero,\bzero}$, and exact charges, 
${S}^{\bzero}_{mn}$, for bulk Ne and SVP, TZVP and QZVP basis sets.}
\label{fig2}
\end{figure}

The total charge associated with the density, $\rho^{\bk,\bq}_{mn}(\br)$, defined in Eq. \ref{Eqn6} is the overlap matrix element 
at wave vector $\bkq$, $S^{\bkq}_{mn}$ (Eq. \ref{Eqn19}). Eqs. \ref{Eqn14} and \ref{Eqn20} were solved for unconstrained coefficients,
$c_\alpha^{\bk,\bq}$, and constrained coefficients, $\overline{c}_\alpha^{\bk,\bq}$ and Lagrange multipliers, $\lambda^{\bk,\bq}$,
for bulk Ne at ($\bk,\bq$) = ($\bzero,\bzero$). Unconstrained coefficients are compared to overlap matrix elements, $S^{\bzero}_{mn}$,
in Fig. \ref{fig2} for DEF2-SVP,-TZVP and -QZVP basis sets \cite{Weigend98} and their corresponding auxiliary basis sets.
Orbital combinations which produce a monopole on site are $<l|l>$ and $<l|l+2n>$ where $l$ is the orbital angular momentum and
$n$ is a positive integer (a product of two 2$p_x$ states or a 2$p_x$ state times 4$f_{x(x^2-3y^2)}$ state on a given site has a 
net monopole, for example). $<s|d>$, $l$ = 0, $n$ = 1 differences are all below $10^{-12}$ and are not shown. 
Absolute differences are shown in Fig. \ref{fig2} and are binned in decades. Distributions are normalized by the number of pairs 
of non-zero overlap matrix elements. Largest differences are in the range 10$^{-1}$ to 10$^{-2}$. Differences do not diminish 
on going from a SVP tp QZVP basis; although the auxiliary basis improves from SVP to QZVP, the number of wave function basis 
functions being fitted increases significantly.

\subsection{Dielectric functions from TDHF}

It is well known that screening of the exchange interaction in the $GW$ approximation, for example, is
necessary to predict the particle-hole gap correctly in solids and that the bare electron-hole attraction term in optical 
excitations in materials is screened in BSE calculations \cite{Rohlfing00}. The $A$ and $B$ matrices in the TDHF method
are equivalent to those in a $GW$/BSE calculation except that: particle-hole energy differences on the diagonal of the $A$ matrix 
are HF eigenvalues in TDHF, rather than $GW$ quasi-particle energies; the electron-hole interaction is the bare interaction in TDHF whereas it
is screened in BSE. In this work, screening of the electron-hole attraction term is sufficiently important, even in wide gap insulators 
such as diamond and MgO, that the experimental spectrum can only be recovered if it is reduced significantly. If it is not
scaled in diamond or MgO the lowest excitation is a strongly localized Frenkel exciton, split off from the continuous part of the dielectric 
function rather than a Wannier exciton with absorption enhancement at lower frequencies in the dielectric function. The former typically 
occurs in systems such as rare gas solids, but not in bulk oxides or semiconductors. 

\begin{table}[!htb] 
\caption{Scale factors, $\alpha$, applied to electron-hole attraction matrix elements, virtual state energy shifts, 
HF band gaps, E$_g^{HF}$, lowest excited states in TDHF (E$^{TDHF}$), band gaps after applying virtual state shifts, E$_g^{scissor}$,
and TDHF lowest excited states with scaled matrix elements and virtual state shifts (E$^{TDHF}_{scaled}$). All energies are in eV. 
Band gaps are direct gaps at $\Gamma$.} %Figures in parentheses are reduction in excitation energies induced by the scaled or unscaled
%electron-hole attraction.} 
\begin{tabular}{lccccccc}
\hline
Material   & $\alpha$ & Shift & E$_g^{HF}$ & E$^{TDHF}$  & E$_g^{scissor}$ & E$^{TDHF}_{scaled}$ \\
\hline                                             
C$^a$      &   0.4    &  7.7  & 14.93      & 12.34       & 7.23            & 6.56 \\                              
MgO        &   0.4    & 11.0  & 19.38      & 14.71       & 8.38            & 7.28 \\
TiO$_2^b$  &   0.4    &  8.2  & 13.71      &  8.64       & 5.51            & 4.12 \\
TiO$_2^c$  &   0.4    &  8.2  & 13.02      &  7.94       & 4.82            & 3.20 \\
%C$^a$      &   0.4    &  7.7  & 14.93      & 12.34(2.59) & 7.23            & 6.56(0.67) \\                              
%MgO        &   0.4    & 11.0  & 19.38      & 14.71(4.68) & 8.38            & 7.28(1.10) \\
%TiO$_2^b$  &   0.4    &  8.2  & 13.71      &  8.64(5.80) & 5.51            & 4.12(1.39) \\
%TiO$_2^c$  &   0.4    &  8.2  & 13.02      &  7.94(5.08) & 4.82            & 3.20(1.62) \\
\hline
$^a$ Diamond \\
$^b$ Anatase \\
$^c$ Rutile \\
\end{tabular}
\label{tab1}
\end{table}

\begin{figure}[h!]
\includegraphics[width=5.0cm,angle=0]{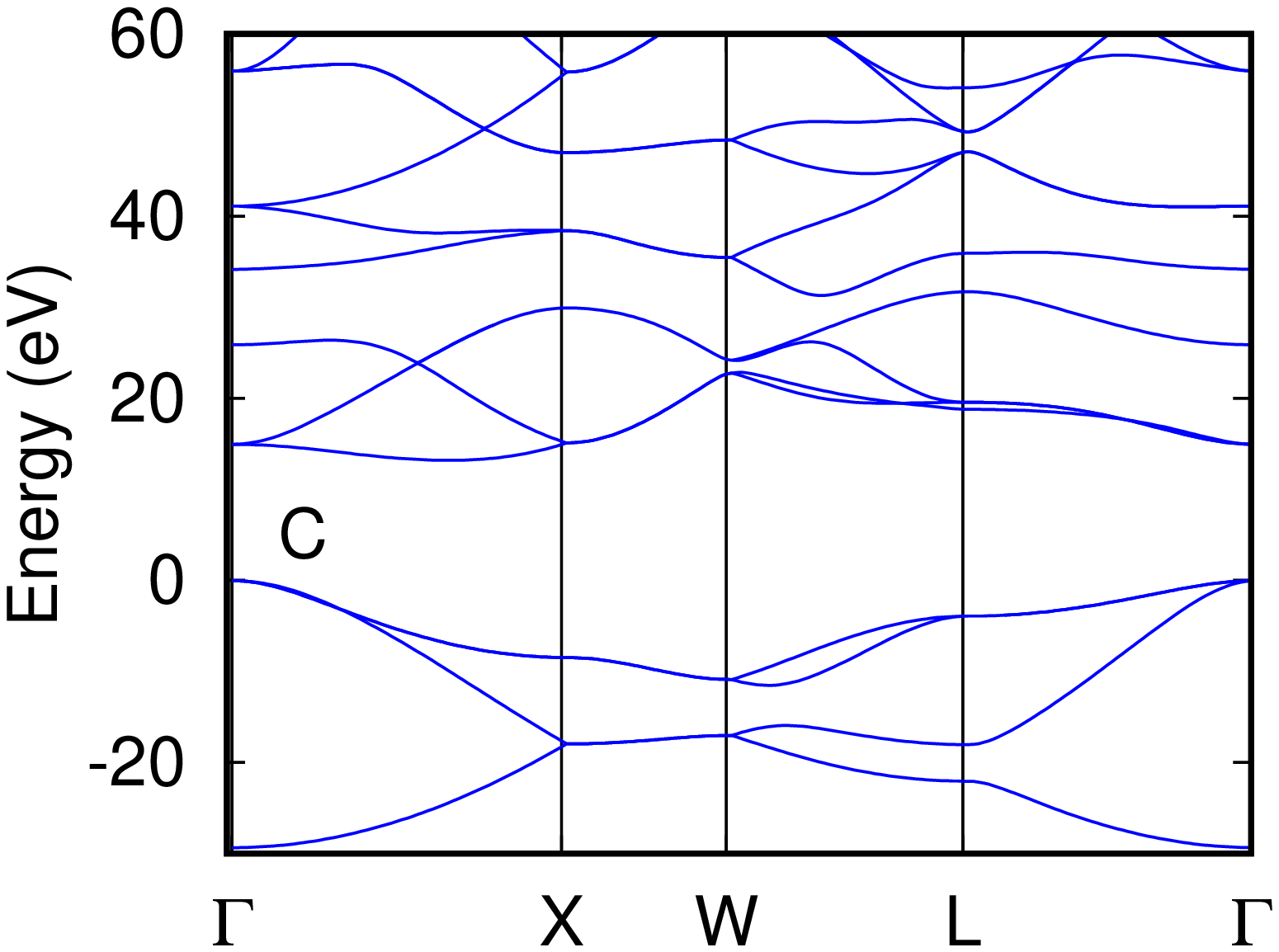}
\includegraphics[width=5.0cm,angle=0]{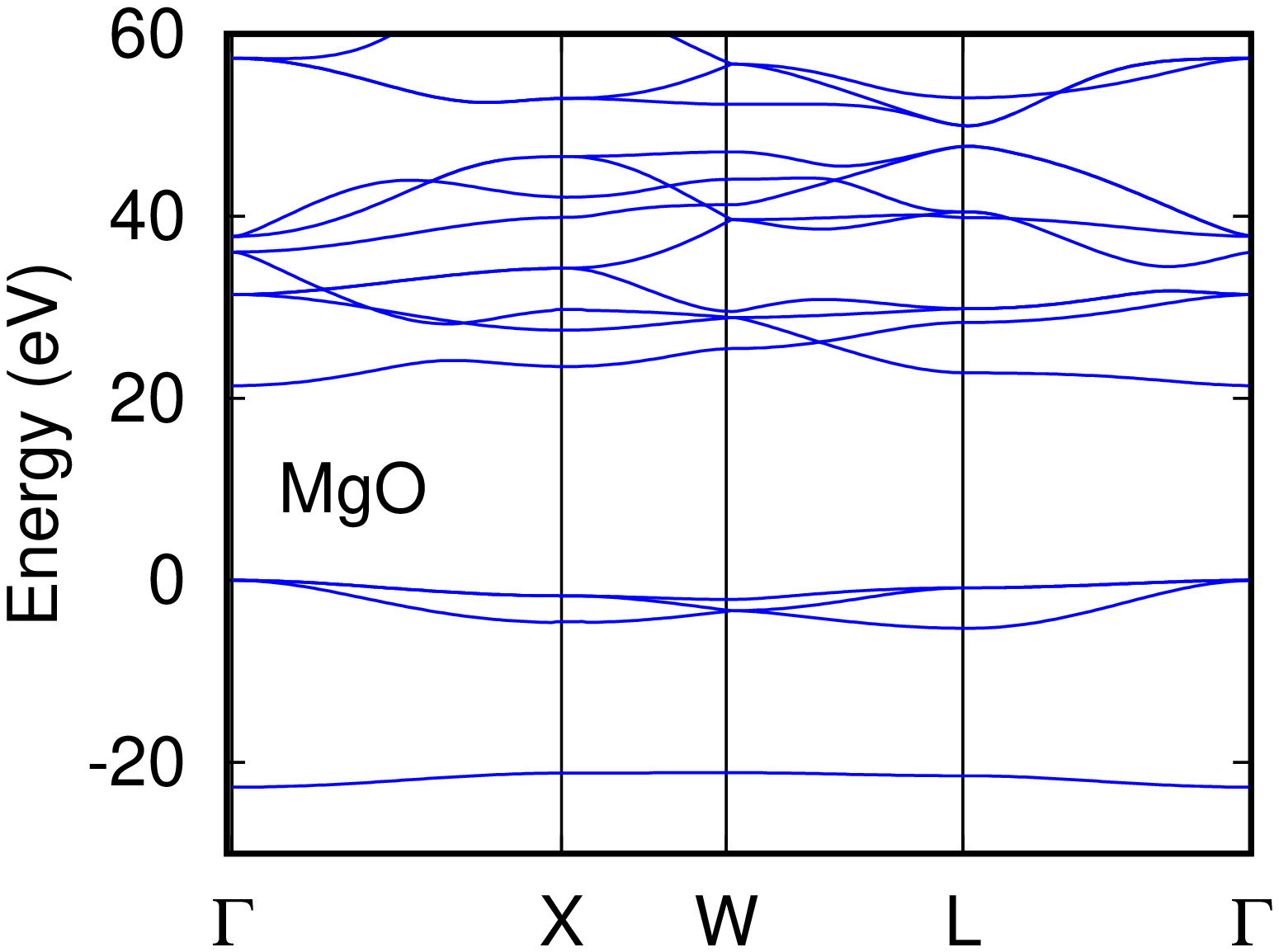}
\includegraphics[width=5.0cm,angle=0]{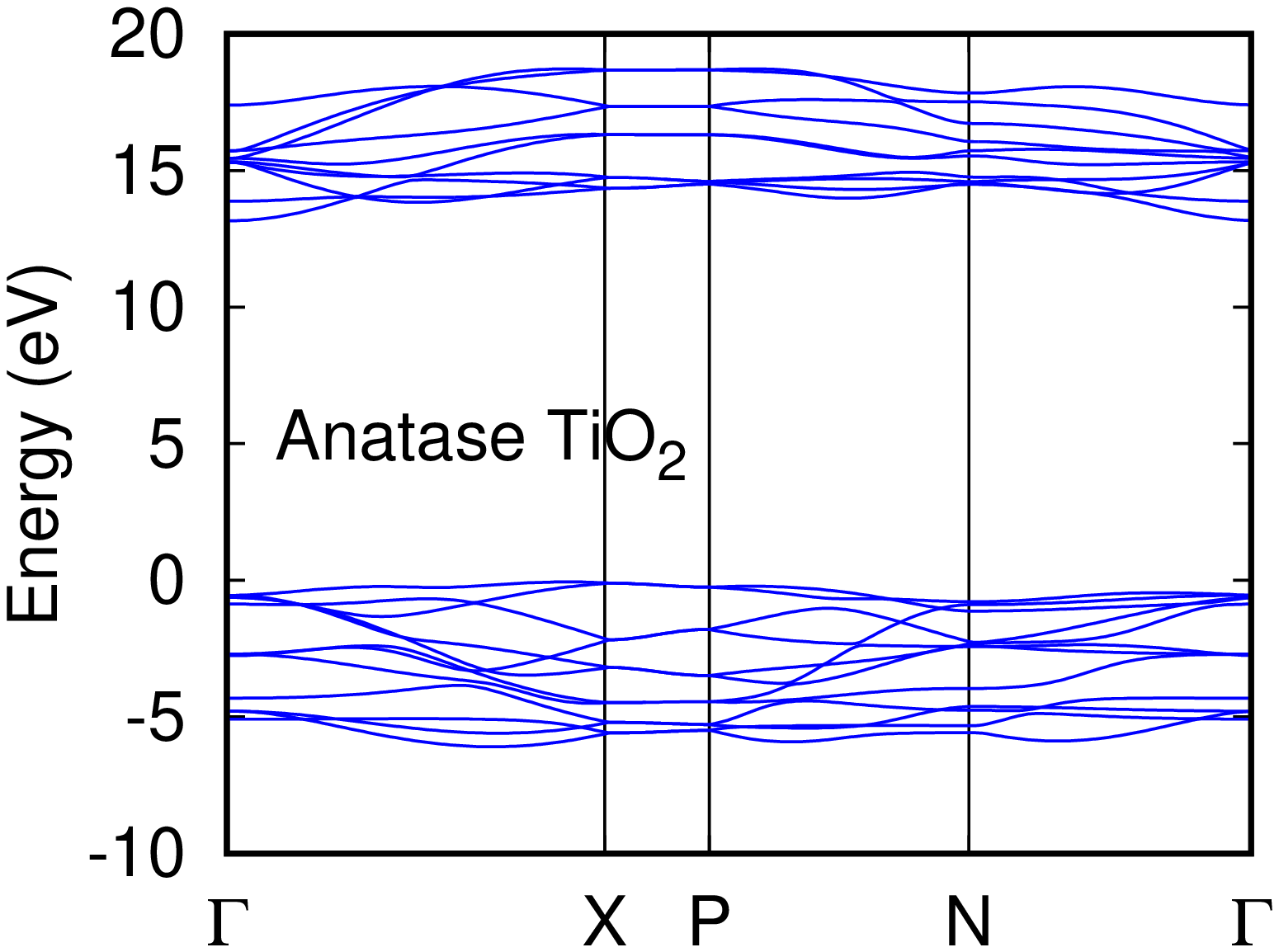}
\includegraphics[width=5.0cm,angle=0]{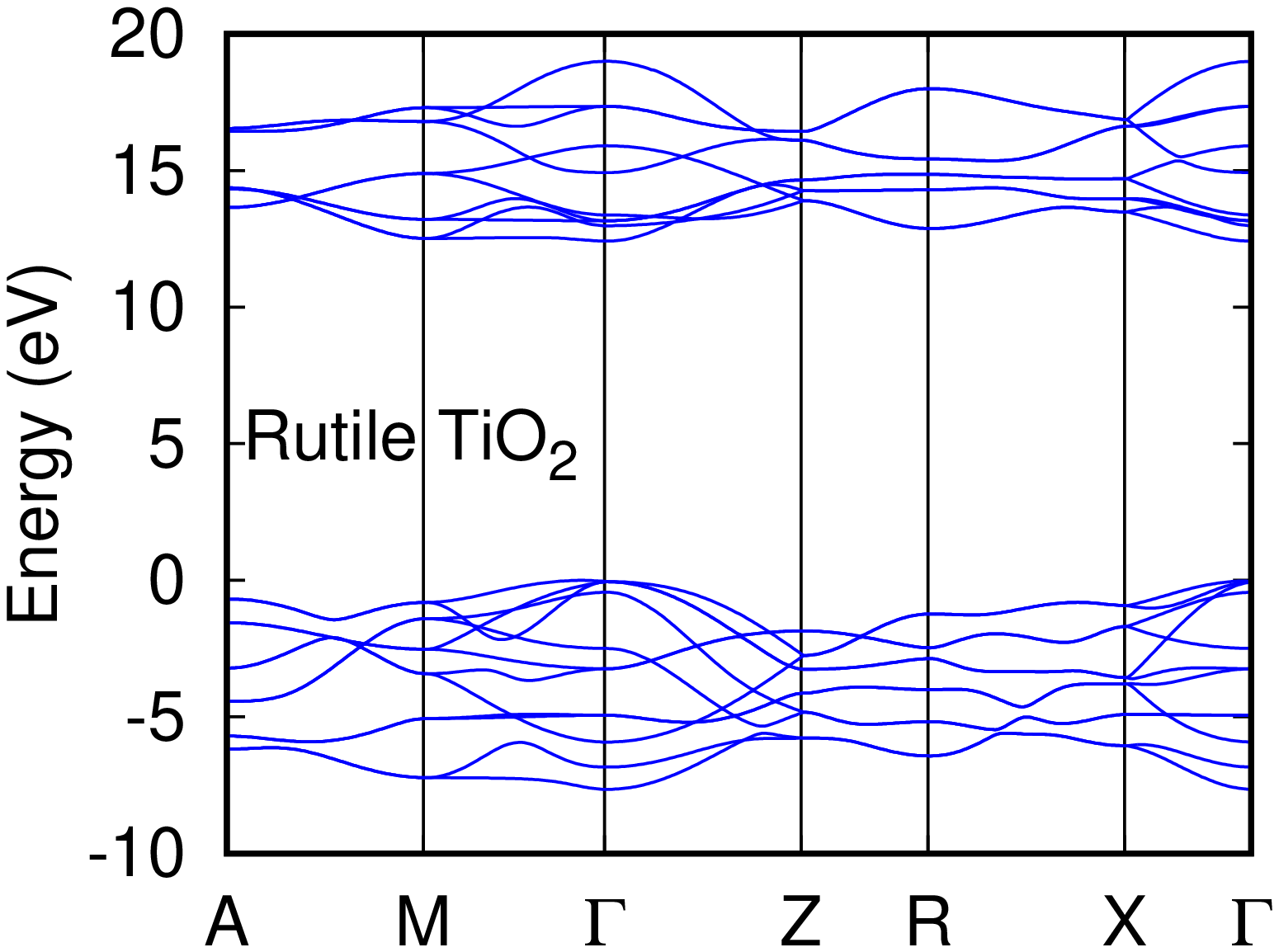}
\caption{(Color online) HF band structures of diamond, rocksalt MgO and anatase and rutile TiO$_2$ polymorphs.}
\label{fig3}
\end{figure}

Electron-hole attraction matrix elements were uniformly scaled by a factor of 0.4 (i.e. with no $\bq$ dependence) for diamond, MgO and 
two polytypes of TiO$_2$. This is similar to the approach of Botti \textit{et al.}, who introduced a term $-\alpha/q^2$ into the TDDFT 
exchange-correlation kernel for a range of insulating materials \cite{Botti04}. In that work, for diamond the optimal value of $\alpha$ was 
0.6 and for MgO it was 1.8. Optimal values for six tetrahedral semiconductors and MgO scaled roughly 
with 1/$\epsilon_\infty$ \cite{Botti04}. Here, a constant scaling by $\alpha$ = 0.4 was found to give satisfactory agreement with experiment
for diamond and the oxides studied, provided a suitable shift of virtual states was used. 
HF band structures of the materials studied in this work are shown in Fig. \ref{fig3} and scaling factors,
$\alpha$, shifts of virtual states and lowest TDHF excitation energies, with and without scaling and shifting, are summarized 
in Table \ref{tab1}. For diamond and MgO spectra were averaged over several calculations with high density sampling in reciprocal space 
($N$ x $N$ x $N$ Monkhorst-Pack nets with $N$ = 10, 11, 12 and 13), whereas for the TiO$_2$ polymorphs a single 6 x 6 x 6 
Monkhorst-Pack net was used. 

\subsubsection{Diamond}

\begin{figure}[h!]
\includegraphics[height=6.0cm,angle=-90]{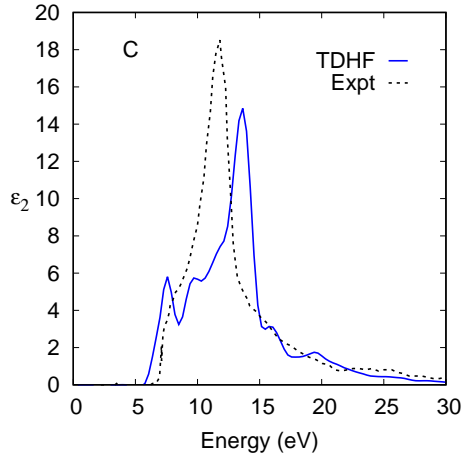}
\caption{(Color online) Imaginary part of the dielectric function of diamond from TDHF calculation and experiment 
(Ref. [\onlinecite{OpticalHandbook}]).}
\label{fig4}
\end{figure}

The TDHF spectrum of diamond obtained after shifting virtual states and scaling electron-hole matrix elements is 
shown in Fig. \ref{fig4}. Four valence and the lowest four conduction band states were used in the TDHF calculation. The overall width 
of spectral features is greater than experiment by around 2.2 eV. The lowest energy excitation in a TDHF calculation with
no shift of virtual states or scaling of the electron-hole attraction matrix elements is 12.34 eV and with scaling described below
it is 6.56 eV.

The position of the main peak 
in the dielectric function occurs at 13.9 eV in the TDHF calculation, while it occurs at 11.8 eV in experiment. The $\Gamma$ point 
HF band gap is 14.93 eV and is combined with a downward shift in virtual states of 7.7 eV, so that the particle-hole gap before the 
TDHF calculation is 7.23 eV. This may be compared with the $G_0W_0$ quasiparticle gap at $\Gamma$ in diamond of 7.5 eV \cite{Hybertsen86}, 
which used a DFT-LDA band structure as input. In that case the width of the spectral features is underestimated and the main peak 
in the dielectric function is underestimated by around 1 eV. The trends of over and underestimating widths of spectral features is 
likely to be the result of similar over and underestimation of valence and conduction band widths by HF and DFT-LDA approximations.
The HF valence band width of diamond has previously been reported to be 28.67 eV \cite{Barnard02} (compared to 29.32 eV in this work), 
versus a DFT-LDA band gap of 21.73 eV \cite{Barnard02}. These values may be compared to the experimental valence band width of diamond of 
23.0 $\pm$ 0.2 eV \cite{Jimenez97} or a $GW$ quasiparticle band width of 23.0 eV \cite{Hybertsen86}.

\subsubsection{Rocksalt MgO}

\begin{figure}[h!]
\includegraphics[height=6.0cm,angle=-90]{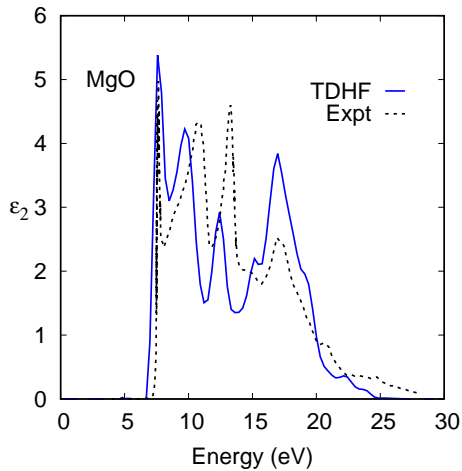}
\caption{(Color online) Imaginary part of the dielectric function of rocksalt MgO from TDHF calculation and experiment 
(Ref. [\onlinecite{Roessler67}]).}
\label{fig5}
\end{figure}

The TDHF dielectric function of MgO after shifting virtual states and scaling matrix elements is shown in Fig. \ref{fig5}. Four valence
and four conduction band states were used in the TDHF calculation. The $\Gamma$ point
HF band gap is 19.38 eV and when combined with a virtual state shift of 11.0 eV, results in a particle-hole gap of 8.38 eV
before the TDHF calculation. This compares with a converged $G_0W_0$ quasi-particle gap of 7.9 eV \cite{Gao16}.
The experimental optical gap of MgO is 7.83 eV \cite{Roessler67,Whited73} and the lowest energy TDHF excitation occurs at 7.59 eV
with the shifted particle-hole gap. If no virtual state shift or matrix element scaling is used, the lowest energy excitation
is a strongly localized exciton at 14.71 eV (Table \ref{tab1}).
The main features of the dielectric function are reproduced in the TDHF calculation, including strong enhancement of optical absorption 
at the threshold energy and three higher energy peaks in the range to 25 eV, which are also reproduced in a BSE calculation \cite{Botti04}.

\begin{figure}[hbp!]
\includegraphics[width=6.0cm,angle=0]{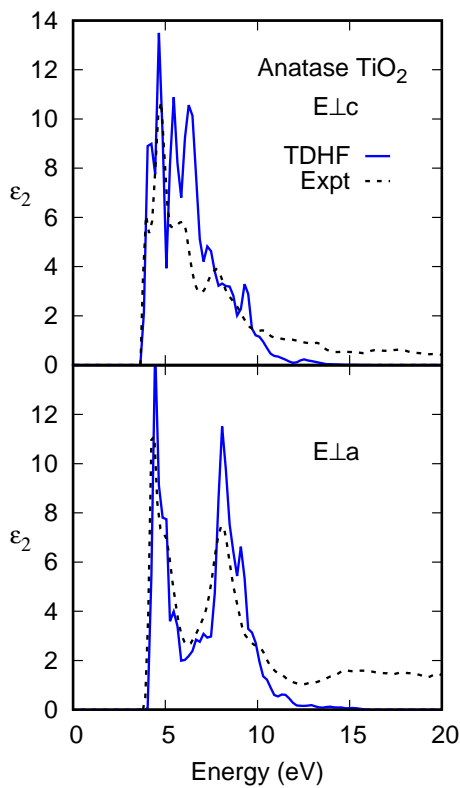}
\caption{(Color online) Imaginary part of the dielectric function of bulk anatase TiO$_2$ from TDHF calculation and experiment 
(Ref. [\onlinecite{Hosaka97}]).}
\label{fig6}
\end{figure}

\begin{figure}[hbp!]
\includegraphics[width=6.0cm,angle=0]{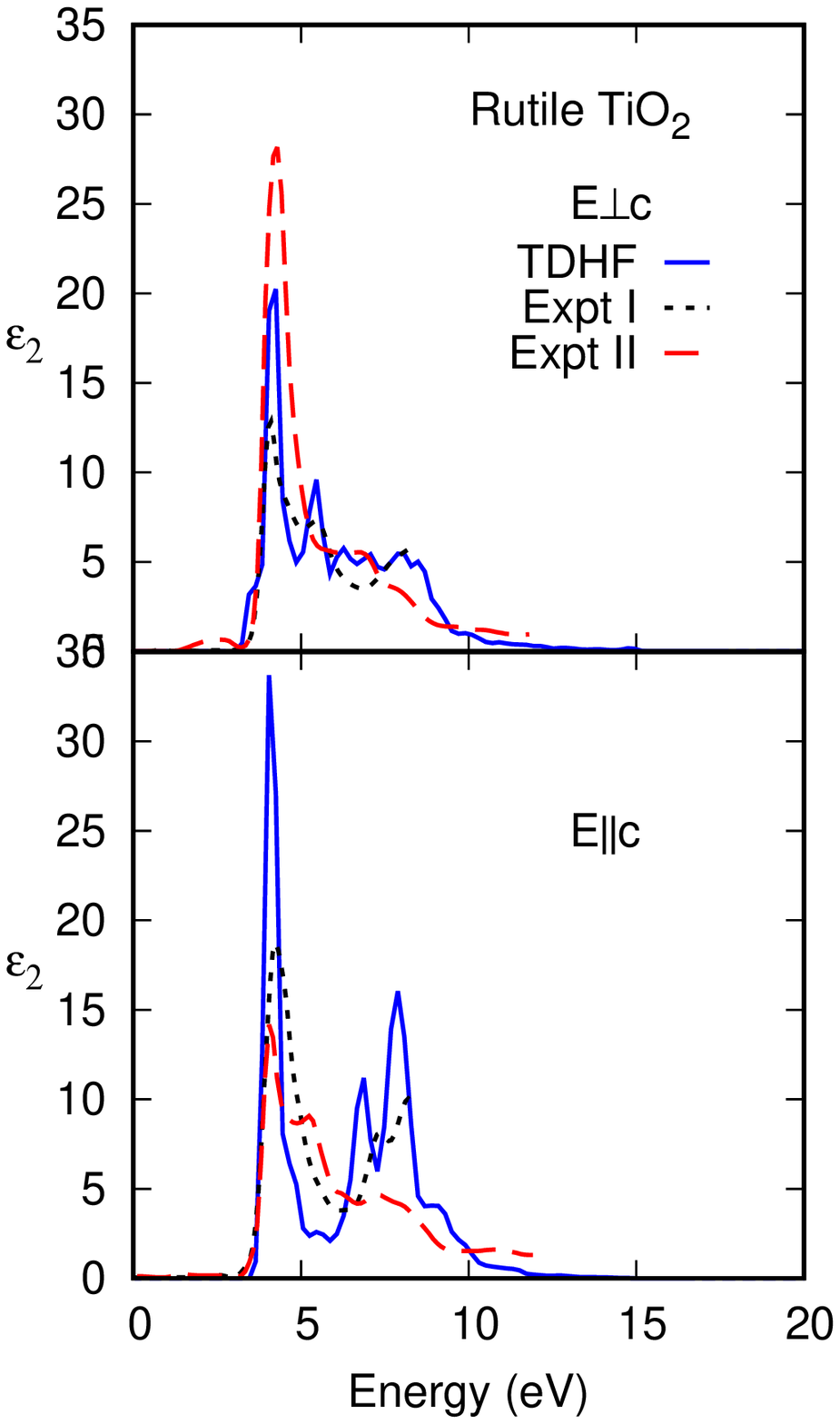}
\caption{(Color online) Imaginary part of the dielectric function of rutile TiO$_2$ from TDHF calculation and experiment 
(Expt I Ref. [\onlinecite{Tiwald00}], Expt II Ref. [\onlinecite{Cardona65}]).}
\label{fig7}
\end{figure}

\subsubsection{Anatase and Rutile TiO$_2$}

The TDHF dielectric functions of anatase and rutile polytypes of TiO$_2$ are compared to experiment in Figs. \ref{fig6} and \ref{fig7}.
Calculations were performed using a 6 x 6 x 6 Monkhorst-Pack net. All 12 predominantly O 2$p$ valence states and 10 predominantly 
Ti 3$d$ conduction band states were included in TDHF calculations and the virtual state shift was 8.2 eV for both polytypes. 

The $\Gamma$ point HF band gap for anatase is 13.71 eV and the indirect gap at the $X$ and $\Gamma$ points is 13.25 eV. 
When combined with a virtual state shift of 8.2 eV, the direct particle-hole gap at $\Gamma$
is 5.51 eV. This may be compared to values of direct $G_0W_0$ quasiparticle gaps at $\Gamma$ for anatase TiO$_2$ of 3.45
\cite{Landmann12}, 4.14 \cite{Kang10} and 4.29 eV \cite{Chiodo10} and indirect gaps of 3.73 \cite{Landmann12}, 3.56 \cite{Kang10} and
3.83 eV \cite{Chiodo10} using plane wave basis sets and a HSE06 hybrid density functional \cite{Landmann12}, DFT-LDA \cite{Kang10} or
DFT-PBE \cite{Chiodo10} density functionals.
For rutile the HF gap is 13.02 eV and the particle-hole gap after the virtual state shift is 4.82 eV. This may be compared to 
values of $G_0W_0$ quasiparticle gaps of 3.46 eV \cite{Landmann12}, 3.38 eV \cite{Kang10} and 3.59 eV \cite{Chiodo10}. 
There is excellent agreement between the TDHF calculations and experiment \cite{Hosaka97} for both polymorphs, in terms of peak position 
and strength. The overall shapes of the spectra with the incident electric vector parallel or perpendicular to the $c$ axis are reproduced.
%there is a shoulder at 4.1 eV in the anatase spectrum with the electric vector perpendicular to the $c$ axis, which has been remarked on in 
%earlier BSE calculations \cite{Lawler08}. The two main peaks in the spectrum with the electric vector perpendicular 
%to the $a$ axis are well reproduced.

Two sets of experimental data have been included in Fig \ref{fig7} for rutile TiO$_2$ \cite{Cardona65,Tiwald00}. They show the low energy 
peak in the experimental dielectric function with the electric vector parallel to the $c$ axis around 4 eV rising to 15 to 18. 
The TDHF calculation shows a somewhat narrower peak rising to over 30. As noted by Landmann \textit{et al.} \cite{Landmann12}, 
the earlier dielectric function measurement for rutile (Ref. [\onlinecite{Cardona65}], Expt II) shows only a shoulder between 5 and 10 eV 
for the electric field vector parallel to the $c$ axis, while TDHF (this work), BSE \cite{Landmann12} and a more recent 
experiment (Ref. [\onlinecite{Tiwald00}], Expt I) \cite{Tiwald00} show a second peak in that range, although these experimental 
data do not extend above 8 eV. This second peak is also found in the BSE calculations mentioned above \cite{Lawler08,Kang10,Landmann12}.

\section{Summary and Conclusions}

A DF method for calculating ring and ladder Coulomb matrix elements over extended Bloch functions has been presented.
Charge densities which arise from products of crystal orbitals at $\bk$ and $\bkq$ are expanded in an auxiliary basis of Gaussian
crystal orbitals. Fitting of these densities requires calculation of two and three center matrix elements of these
crystal orbitals over a lattice modulated Ewald potential. In a robust fit of the densities, expansion coefficients are 
obtained by inverting the two center matrix elements. Alternatively, a variational fit of these densities, in which the 
integrated charge densities are constrained to have their exact values, requires solution of a set of linear equations 
with Lagrange multipliers.

We have demonstrated that the DF method described is able to recover Coulomb and exchange energies for
several periodic systems to a similar level of accuracy as has been reported elsewhere for periodic systems 
\cite{Milko07,Burow09,Sun17a}. For light atoms such as C, O or Mg the SCF Coulomb energy is reproduced to within 50 $\mu$H per atom,
including core contributions. The exchange energy extrapolated to infinite sampling density agrees with the SCF exchange energy
to within 1 mH per atom.

TDHF calculations presented here use uniform scaling of the electron-hole attraction matrix elements rather than a $\bq$-dependent
inverse dielectric matrix and shift the HF virtual states downward by a constant amount in order to achieve agreement with experiment
for diamond and three oxide compounds. The shifted virtual state band gaps used are similar to $G_0W_0$ band gaps, but may exceed the
$G_0W_0$ band gap. There may therefore be a range of scaling of the electron-hole attraction
matrix elements and virtual state shift which yield comparable agreement with experiment. For MgO, a particle-hole gap 
around 0.4 eV greater than the $G_0W_0$ gap and gave good agreement with experiment for the onset of optical
absorption. For anatase and rutile TiO$_2$, values around 1.5 eV greater than $G_0W_0$ values were used and good agreement with onset of
optical absorption was predicted.

\section{Data availability}

The data that support the findings of this study are available from the corresponding author upon reasonable request.

\begin{acknowledgements}
The author gratefully acknowledges helpful discussions with R. Dovesi, L. Maschio and S. Trickey.
Calculations were performed on the Kelvin and Boyle clusters maintained by the Trinity Centre for High Performance Computing,
funded through grants from Science Foundation Ireland and the Irish Higher Education Authority. 
\end{acknowledgements}

\appendix

\section{Derivation of Eq. 10}

Equations \ref{Eqn12} and \ref{Eqn13} are derived for a product of CO with wave vectors $\bk$ and $\bkq$, 

\begin{equation}
\label{EqnA1}
\rho^{\bk,\bq}_{mn}(\br) = \sum_{\bA,\bB}\phi_m^*(\br-\bA)\phi_n(\br-\bB) e^{-i\bk.\bA + i(\bkq).\bB},
%\rho^{\bq}_{mn}(\br) = \sum_{\bA,\bB}\phi_m^*(\br-\bA)\phi_n(\br-\bB) e^{-i\bk.\bA + i(\bkq).\bB},
\end{equation}

\noindent
The matrix element in Eq. \ref{Eqn4} contains two factors which are products of CO with wave vectors, $\bk$ and $\bkq$,
and where the second factor is the conjugate of the first. The charge densities for which the fitting errors are minimized are
therefore, $\rho^{\bk,\bq}_{mn}(\br)$ and the conjugate, $\rho^{*\bk,\bq}_{rs}(\br)$. Hence fitting densities,
%therefore, $\rho^{\bq}_{mn}(\br)$ and the conjugate, $\rho^{*\bq}_{rs}(\br)$. Hence fitting densities,

\begin{equation}
\label{EqnA2}
\tilde{\rho}^{\bk,\bq}_{mn}(\br)   = c_\alpha^{\bk,\bq} \sum_{\bA} \chi_\alpha (\br-\bA)e^{i\bq.\bA}
%\tilde{\rho}^{\bq}_{mn}(\br)   = c_\alpha \sum_{\bA} \chi_\alpha (\br-\bA)e^{i\bq.\bA}
\end{equation}

\noindent
and

\begin{equation}
\label{EqnA3}
\tilde{\rho}^{*\bk,\bq}_{rs}(\brp) = c^{*\bk,\bq}_\beta \sum_{bB} \chi^{*}_\beta (\brp-\bB)e^{-i\bq.\bB}
%\tilde{\rho}^{*\bq}_{rs}(\brp) = c^*_\beta \sum_{bB} \chi^{*}_\beta (\brp-\bB)e^{-i\bq.\bB}
\end{equation}

\noindent
are used.

\begin{widetext}

Minimization of an integral of the form in Eq. \ref{Eqn2}, using the Coulomb metric, by applying $\frac{\partial}{\partial c^*_\beta}$ yields,

\begin{equation}
\label{EqnA4}
\sum_{\bB} \drdrp \frac{\rho^{\bk,\bq}_{mn}(\br) \chi^*_\beta(\brp-\bB)}{|\br - \brp|} = c^{\bk,\bq}_\alpha \sum_{\bA,\bB}\drdrp \frac{\chi_\alpha (\br-\bA) \chi^*_\beta(\brp-\bB)}{|\br-\brp|} e^{i\bq.(\bA-\bB)},
%\sum_{\bB} \drdrp \frac{\rho^{\bq}_{mn}(\br) \chi^*_\beta(\brp-\bB)}{|\br - \brp|} = c_\alpha \sum_{\bA,\bB}\drdrp \frac{\chi_\alpha (\br-\bA) \chi^*_\beta(\brp-\bB)}{|\br-\brp|} e^{i\bq.(\bA-\bB)},
\end{equation}

\end{widetext}

\noindent
and a similar equation for $c^{*\bk,\bq}_\beta$ by applying $\frac{\partial}{\partial c_\alpha}$. Making the substitutions, 
$\bB = \bA + \bB'$, $\brp - \bB' = \brpp$ and using lattice translational invariance, the right hand side becomes,

\begin{equation}
\label{EqnA5}
c^{\bk,\bq}_\alpha \sum_{\bB'} \drdrp \frac{\chi_\alpha (\br) \chi^*_\beta(\brpp)}{|\br-\brpp-\bB'|} e^{-i\bq.\bB'} = c^{\bk,\bq}_\alpha V^{\bq}_{\alpha\beta}
%c_\alpha \sum_{\bB'} \drdrp \frac{\chi_\alpha (\br) \chi^*_\beta(\brpp)}{|\br-\brpp-\bB'|} e^{-i\bq.\bB'} = c_\alpha V^{\bq}_{\alpha\beta}
\end{equation}

\noindent
Making the substitutions, $\bB = \bA + \bA'$, $\bC = \bA + \bC'$ and $\brp - \bC' = \brpp$ and using lattice translational invariance,
the left hand side becomes,

\begin{equation}
\label{EqnA6}
\sum_{\bA',\bC'} \drdrp \frac{\phi_m^*(\br)\phi_n(\br-\bA') \chi^*_\beta(\brpp)}{|\br-\brpp-\bC'|} e^{i(\bk+\bq).\bA' -i\bq.\bC'} = V^{\bk,\bq}_{mn\beta},
%\sum_{\bA',\bC'} \drdrp \frac{\phi_m^*(\br)\phi_n(\br-\bA') \chi^*_\beta(\brpp)}{|\br-\brpp-\bC'|} e^{i(\bk+\bq).\bA' -i\bq.\bC'} = V^{\bq}_{mn\beta},
\end{equation}

\noindent
as given in Eqn. \ref{Eqn12} and \ref{Eqn13}.

\section{Derivation of Eq. 19}

The constraint on charge densities of orbital products in Eq. \ref{Eqn18} may be expressed as,

\begin{equation}
\label{EqnB1}
\dr (\rho^{\bk,\bq}_{mn}(\br) - \overline{c}^{\bk,\bq}_{\alpha}\chi^{\bq}_\alpha(\br)) = 0,
%\dr (\rho^{\bq}_{mn}(\br) - \overline{c}^{\bq}_{\alpha}\chi^{\bq}_\alpha(\br)) = 0,
\end{equation}

\noindent
%Substituting for $\rho^{\bq}_{mn}(\br)$ and $\chi^{\bq}_\alpha(\br)$ using Eq. \ref{Eqn6} and \ref{Eqn7} and replacing the
Substituting for $\rho^{\bk,\bq}_{mn}(\br)$ and $\chi^{\bq}_\alpha(\br)$ using Eq. \ref{Eqn6} and \ref{Eqn7} and replacing the
lattice vector $\bB$ by $\bA+\bB'$ yields, 

\begin{equation}
\label{EqnB2}
\sum_{\bB'}\dr\left(\phi_m^*(\br)\phi_n(\br-\bB') e^{i(\bkq).\bB'} - \overline{c}^{\bk,\bq}_{\alpha} \chi_\alpha(\br) \right) = 0,
\end{equation}

\noindent
or,

\begin{equation}
\label{EqnB3}
S_{mn}^{\bkq} = \overline{c}^{\bk,\bq}_{\alpha} \left< \chi_\alpha \right>,
\end{equation}

\noindent
where,

\begin{equation}
\label{EqnB4}
S_{mn}^{\bkq} = \sum_{\bB'} \dr \phi_m^*(\br)\phi_n(\br-\bB') e^{i(\bkq).\bB'}
\end{equation}

\noindent
and,

\begin{equation}
\label{EqnB5}
\left< \chi_\alpha \right> = \dr \chi_\alpha(\br)
\end{equation}

\bibliography{paper}

\end{document}